\documentclass[11pt]{article}

\usepackage{graphicx,latexsym}

\begin{document}

\begin{flushright}
hep-ph/0307275\\
KIAS-P03056\\
\end{flushright}

\begin{center}
{\Large {\bf Signatures of quantized TeV scale Black holes in
scattering processes}}\\
\hspace{10pt}\\
S. R. Choudhury\footnote{electronic address: src@physics.du.ac.in}\\
{\em Department of Physics, Delhi University, Delhi 110007,
India}\\
\hspace{10pt}\\
A. S. Cornell\footnote{electronic address: alanc@kias.re.kr}\\
{\em Korea Institute for Advanced Study, 207-43}\\
{\em Cheongryangri 2-dong, Dongdaemun-gu, Seoul 130-722}\\
\hspace{10pt}\\
G. C. Joshi\footnote{electronic address:
joshi@physics.unimelb.edu.au} and B. H. J.
McKellar\footnote{electronic address:
b.mckellar@physics.unimelb.edu.au}\\
{\em School of Physics, University of Melbourne,}\\
{\em Parkville, Victoria 3010, Australia}\\
\hspace{10pt}\\
$22^{nd}$ of July, 2003.
\end{center}

\begin{abstract}
In this paper we shall study the phenomenology of a doubly charged
and neutral exchange black hole in an $(n+3)$ extra-dimensional
scenario, where the black hole shall be treated a normal quantum
field.
\end{abstract}

\section{Introduction}
\hspace{0.5cm} Extra-dimensional scenarios have been put forward
in recent times as a possible solution to the so called hierarchy
problem \cite{Extra}.  Such models typically bring down the Planck
scale from their traditional high values to a scale in the TeV
region. While there would be correctional terms to measurables in
high energy processes coming from excitations of standard model
particles, there is also the very interesting possibility that
such models would give rise to the presence of black holes in the
TeV region rather than the normal Planck mass scale \cite{Banks}.
Since TeV scale laboratory collisions are well within view, we
should expect, if these models are valid, black hole production in
the laboratory.  The detection of such black holes would be
usually studied by its decay products in such collisions.

\par In the classical  scenario \cite{Hawking}, one would
expect a black hole  to radiate through a black body radiation
spectrum. This would be reflected in the decay products having
large multiplicity and large transverse energy
\cite{Dimopoulos:2001hw}. One of course has no clue at the moment
whether such classical ideas would survive in a full quantum
theory of  gravitation which of course does not exist till now.
However, this picture cannot survive if the black hole mass is
close to the Planck mass. As the black hole starts radiating its
mass would decrease and approach the Planck mass, where Hawking's
approximation breaks down. The radiation emitted at that stage is
probably highly non-thermal \cite{Witten}.

\par The situation becomes more reminiscent of atomic radiation if one
accepts some current ideas regarding quantization of black hole
mass.  This idea of quantization was first put forward by
Bekenstein \cite{Bekenstein} and has been dealt with by a number
of authors since.  If the quantization picture is correct then the
ideas mentioned above, regarding radiation emitted by a black hole
according to the black body radiation formula, may be expected via
the correspondence principle only when the level splitting becomes
very small compared with the mass.  The ground state of such a
system would be stable against such radiation but would probably
decay into ordinary particles. If the Planck mass is indeed in the
TeV region, then  as we go to higher and higher values of energy
in the laboratory, the ground state black hole would be the first
one to be reached in a quantized black hole scenario.  It is thus
interesting then to think about the possible signatures for that.

\par In a recent publication Bilke et al. \cite{Bilke} have put
forward the interesting idea that for such quantized black holes,
for low values of quantum numbers, the black hole should be
treated like any other particle, describable by a quantum field
which interacts with other standard model fields.  They
concentrate on a doubly charged black hole and work out possible
signatures of that in a possible electron-electron TeV scale
scattering.  We find this idea very interesting and the purpose of
this note is to further consider this picture in relation to
neutral systems.

\par The quantization formula for the black hole mass has the form
\begin{equation}
M_{bh}^2  =  g  M_P^2 \left[n \left(1 + \alpha \frac{q^2}{2 n}
\right) + \frac{j^2}{n} \right] \label{eq1}
\end{equation}
where $M_P$ is the Planck mass, $\alpha$ the fine structure
constant, $n$ the radial quantum number, $j$ the angular momentum
of the black hole as constrained by
$$ j^2 + \frac{\alpha q^2}{4} \leq n^2 .$$
$g$ is a pre-factor whose actual value is somewhat uncertain.  We
shall take the value of $g$ as given by Khriplovich \cite{khr} and
we shall see that the type of conclusions we wish to draw are
independent of this particular choice. The central point about the
relevance of this mass formula for high energy scattering is that
it not only relates the masses and the ground state, and its
excitations, but also predicts that charged and neutral black
holes must exist together with unequal but correlated masses. This
is completely different from the situation in a hadronic resonance
where charged and neutral resonant states have only
electromagnetic splittings.  Not only that, as we show below, the
neutral and charged black holes in the ground state will exhibit
different characteristics in high energy collisions.

\par Consider first a doubly charged black hole in the ground state $n=1$
which we represent as discussed above by a scalar field $\phi_c$.
Its interaction with electron field $\psi$ can be written down as
\begin{equation}
L^c_{int}  =  \frac{i \kappa^c}{2} M^c_{bh} \phi_c \cdot
\tilde{\psi} C \psi \label{eq2}
\end{equation}
where $\kappa^c$ is the effective coupling constant,$C$ is the
charge conjugation matrix and $M^c_{bh}$ is the mass of the doubly
charged black hole with $q=2$ and $j=0$.  As shown in by Bilke et
al. \cite{Bilke}, the coupling constant $\kappa^c$ can be related
to the Schwarzschild radius $R_s$ in $(n+3)$-dimensions
\cite{Myers}:
\begin{eqnarray}
\kappa^c  &=&  2  R_s\\
R_s &=& \frac{1}{\sqrt{\pi} M_P} \left[ \frac{M^c_{bh}}{M_P}
\left( \frac{8 \Gamma(\frac{n+3}{2})}{n+2} \right)
\right]^{\left(\frac{1}{n+1} \right)} .
\end{eqnarray}
The matrix element for the scattering process
$$ e^-(p_1) + e^-(p_2) \to  e^-(p_3)  + e^- ( p_4) $$
through the black hole of mass $M^c_{bh}$ considered as a
Breit-Wigner resonance of width $\Gamma^c_{tot}$ is
\begin{eqnarray}
M(p_1,p_2;p_3,p_4) &=& M^{\star}(p_3,p_4) \left\{ \frac{1}{s - (
M^c_{bh}+  \frac{i \Gamma^c_{tot}}{2} )^2} \right\}   M(p_1,p_2)\\
M(p,q) &=& i \kappa^c M^c_{bh} \tilde{u}(p) C u(q)
\end{eqnarray}
where $s$ is the centre of mass energy squared. We can readily
calculate the total cross section with this amplitude. Neglecting
masses in the incoming and outgoing channels, we get
\begin{equation}
\sigma(s) = \frac{(\kappa^c M^c_{bh})^4 s}{16 \pi} \frac{1}{|s -
(M^c_{bh}+ \frac{i \Gamma^c_{bh}}{2})^2 |^2 } \label{eq7}
\end{equation}
The cross section calculated above differs from the one calculated
by Bilke et al. \cite{Bilke} in one essential respect.  Unlike
them we have not assumed that the resonant black hole emits
particles with a black body radiation spectrum which would have
necessitated multiplying the phase space at every point with a
statistical factor. We feel justified in doing so since at the
ground level, there is no classical Hawking radiation and
therefore there is no question of particles being emitted as if
they are emitted from a black body of a certain temperature.  As
emphasized earlier, the hypothesis we are testing is whether a
black hole after emitting all the Hawking type of radiation  it
could and falling to the ground state can be treated like any
other particle describable by a field.

\par The cross section given above as such is not suitable for
direct comparison with experimental data since it involves an
unknown quantity, namely the total width.  The elastic width
$\Gamma^c$ can of course be easily calculated from the basic
interaction given above and turns out to be
\begin{equation}
\Gamma^c = \frac{(\kappa^c)^2  (M^c_{bh})^3}{8 \pi} .
\end{equation}
Bilke et al. \cite{Bilke} make the assumption that the total and
the elastic widths are equal.  We believe that this assumption is
justified in the charged case but not in the neutral case and this
is the key to the difference in behaviour of the two.  To show
this, consider an extra particle (considered spinless for
simplicity) described by a field $\chi$ produced in the collision
along with the electrons.  The coupling of the charged black hole
to the three particle state of two electrons and the scalar is
effectively described by the interaction
\begin{equation}
L^{\chi} = \frac{i \kappa^c}{2 M_P} M^c_{bh} \phi_c \chi \cdot
\tilde{\psi} C \psi ,
\end{equation}
where the Planck mass in the denominator effectively compensates
the extra dimension introduced by the scalar field in the
Lagrangian.  There is of course no proof that the effective
Lagrangian above should be related to the Lagrangian given in
equation (\ref{eq2}) except on dimensional grounds, and the fact
we are dealing with interactions at scales defined by the Planck
mass. With this Lagrangian we can easily calculate the cross
section for the process
$$ e^- +  e^-  \to  e^-  +  e^-  +  \chi $$
and we get
\begin{equation}
\sigma( e^- + e^- \to e^- + e^- + \chi ) = \left[ \frac{s}{48
\pi^2 M_P^2} \right] \sigma( e^- + e^- \to e^- + e^-) .
\end{equation}
The factor in square brackets on the right hand-side is of course
much less than one.  This situation will become worse for
multi-particle production and hence we find that the assumption
made by Bilke et. al \cite{Bilke} is justified.

\par Consider now an electrically neutral process, like a pair of
photons, producing a black hole. Like the charged case, the
interaction can be defined by a Lagrangian:
\begin{equation}
L_{int} = \frac{\kappa}{M_{bh}}  \phi . F_{\mu\nu} F^{\mu\nu} ,
\end{equation}
where $\phi$ represents the neutral black hole field, $M_{bh}$ its
mass given be equation (\ref{eq1}) and $\kappa$ represents the
coupling.  Just as in the charged case, we can relate $\kappa$ to
the Schwarzschild radius $R_s$ and we get an identical result
$\kappa = 2 R_s$.  The calculation for the cross section
$$ \gamma + \gamma \to \gamma  + \gamma  $$
through the neutral black hole proceeds exactly parallel to the
charged case with identical results.  The crucial difference comes
about in this case   because although multi-particle production
will be suppressed as in the charged case there are other two
particle channels available for coupling to the neutral black
hole. Neglecting particle masses, all these two particle channels
will be produced approximately with the same cross section  and
hence in this the total width is no longer going to be equal to
the elastic one but will be the elastic one multiplied by the
number of two particle channels.  In the standard model it is
easily seen there will be 28 open two body channels.  The cross
section for the process
$$ \gamma  +  \gamma  \to  \gamma + \gamma $$
will again be given by equation (\ref{eq7}) with this
understanding of the total cross section. The total cross section
in this case of course will be 28 times the elastic one. This
difference between the neutral and charged cross section as
outlined above dictates that the neutral one should be much more
flatter  with the peak occurring at a slightly lower value, as
predicted by the mass formula equation (\ref{eq1}).

\par Figure 1 shows our result for a value of the Planck mass
equal to 1TeV and the pre-factor $g$ taken to be $g= 0.614/\pi$.
The exact position of the peak of course depends on $g$ which as
stated before has some uncertainty, but the relative position of
the peak is independent of $g$.  In figure 2 we show the same
results for different values of the number of extra dimensions
$n$.

\par In conclusion, we have shown that in the extra-dimensional
scenario with quantized masses of the black hole, there are
specific signatures which can be examined in a TeV scale
scattering of particles.  The chief characteristics are the
necessary occurrence of a peak for neutral processes at a mass
lower by a known factor from the peak position in a doubly charged
scattering process.  Further, characteristically, the neutral peak
may be expected to much flatter as compared to the charged one.
All these seem to be realizable experimentally in the foreseeable
future.

\section*{Acknowledgements}
SRC would like to acknowledge the Department of Science and
Technology, Government of India for support through a research
grant.  This work was supported by the Australian Research Council
and a grant from Melbourne University.

\pagebreak

\begin{figure}
\begin{center}
\includegraphics[angle=270,width=10cm]{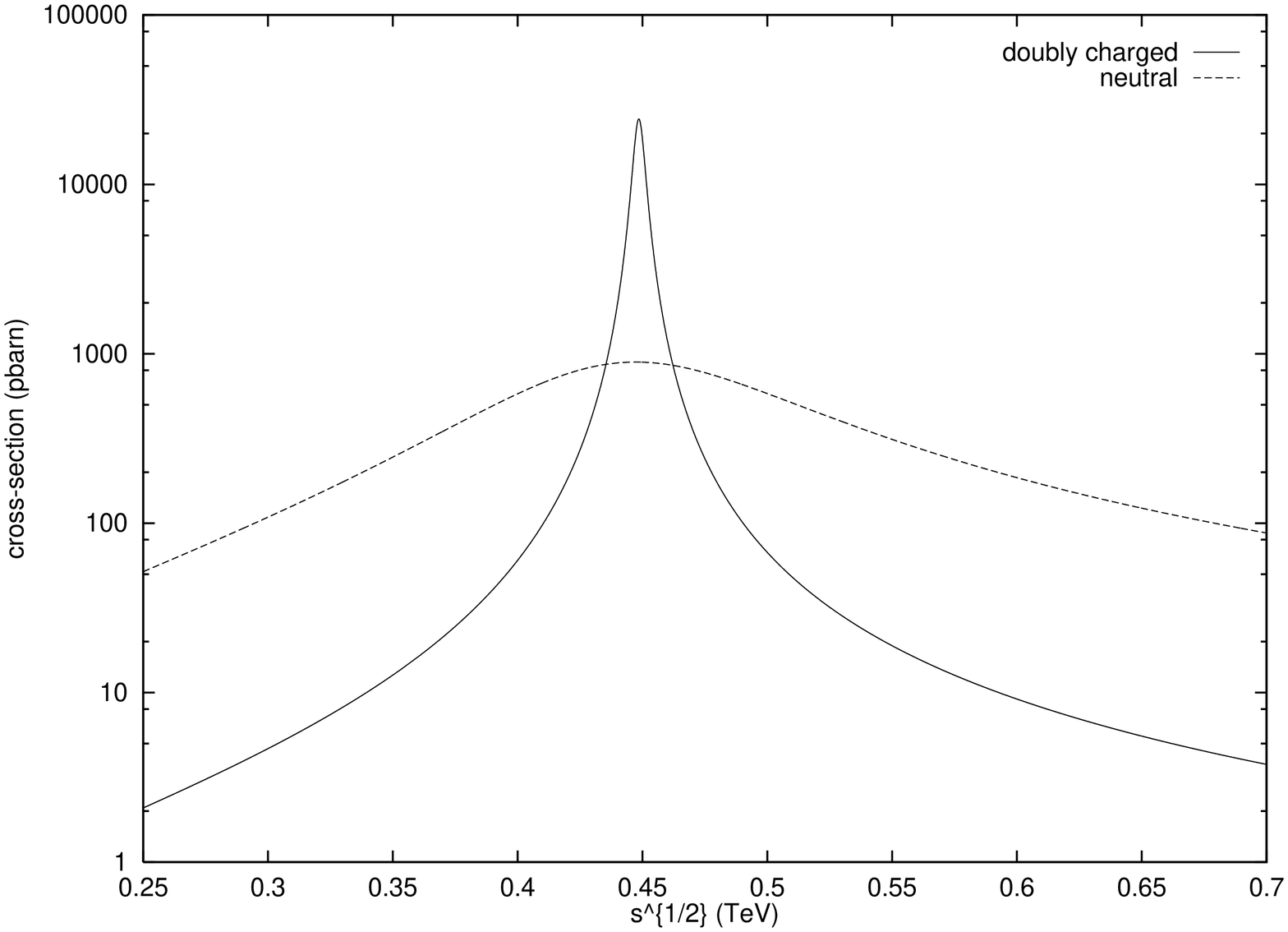}
\caption{The total cross sections in pb for the doubly charged and
neutral channels.  $M_P$ has been takes as 1TeV, whereas the
pre-factor $g$ has been taken as $0.614/\pi$.}\label{fig1}
\end{center}
\end{figure}

\begin{figure}
\begin{center}
\includegraphics[angle=270,width=10cm]{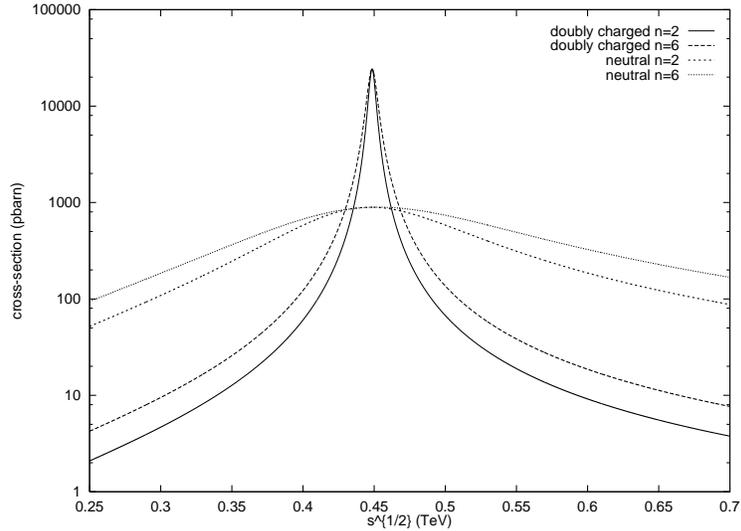}
\caption{The total cross sections in pb for the doubly charged and
neutral channels where the number of extra-dimensions $n$ has been
taken to be 2 or 6. $M_P=1TeV$ and the pre-factor
$g=0.614/\pi$.}\label{fig2}
\end{center}
\end{figure}


\begin{thebibliography}{2}

\bibitem{Extra}
N.~Arkani-Hamed, S.~Dimopoulos and G.~R.~Dvali,
Phys.\ Lett.\ B {\bf 429}, 263 (1998) [arXiv:hep-ph/9803315];
I.~Antoniadis, N.~Arkani-Hamed, S.~Dimopoulos and G.~R.~Dvali,
Phys.\ Lett.\ B {\bf 436}, 257 (1998) [arXiv:hep-ph/9804398];
N.~Arkani-Hamed, S.~Dimopoulos and G.~R.~Dvali,
Phys.\ Rev.\ D {\bf 59}, 086004 (1999) [arXiv:hep-ph/9807344];
L.~Randall and R.~Sundrum,
Phys.\ Rev.\ Lett.\  {\bf 83}, 3370 (1999) [arXiv:hep-ph/9905221].

\bibitem{Banks}
T.~Banks and W.~Fischler,
arXiv:hep-th/9906038;
S.~B.~Giddings and S.~Thomas,
Phys.\ Rev.\ D {\bf 65}, 056010 (2002) [arXiv:hep-ph/0106219].

\bibitem{Hawking}
S.~W.~Hawking,
Commun.\ Math.\ Phys.\  {\bf 43}, 199 (1975);
J.~B.~Hartle and S.~W.~Hawking,
Phys.\ Rev.\ D {\bf 13}, 2188 (1976);
J.~M.~Bardeen, B.~Carter and S.~W.~Hawking,
Commun.\ Math.\ Phys.\  {\bf 31}, 161 (1973).
S.~W.~Hawking,
Phys.\ Rev.\ D {\bf 13}, 191 (1976);
G.~W.~Gibbons and S.~W.~Hawking,
Phys.\ Rev.\ D {\bf 15}, 2738 (1977).

\bibitem{Dimopoulos:2001hw}
S.~Dimopoulos and G.~Landsberg,
Phys.\ Rev.\ Lett.\  {\bf 87}, 161602 (2001)
[arXiv:hep-ph/0106295];
S.~Dimopoulos and R.~Emparan,
Phys.\ Lett.\ B {\bf 526}, 393 (2002) [arXiv:hep-ph/0108060];
M.~Cavaglia,
arXiv:hep-ph/0305256.

\bibitem{Witten}
E.~Witten,
Nucl.\ Phys.\ Proc.\ Suppl.\  {\bf 91}, 3 (2000)
[arXiv:hep-ph/0006332].

\bibitem{Bekenstein}
J.~D.~Bekenstein,
Lett.\ Nuovo Cim.\  {\bf 11}, 467 (1974).

\bibitem{Bilke}
S.~Bilke, E.~Lipartia and M.~Maul,
arXiv:hep-ph/0204040.

\bibitem{khr}
I.~B.~Khriplovich,
Phys.\ Lett.\ B {\bf 537}, 125 (2002) [arXiv:gr-qc/0109092].

\bibitem{Myers}
R.~C.~Myers and M.~J.~Perry,
Annals Phys.\  {\bf 172}, 304 (1986).

\end{thebibliography}
\end{document}